\documentclass[useAMS,usenatbib]{mn2e}
\usepackage{ulem}
\usepackage[dvips]{graphicx}

\title[Measuring the dark matter equation of state]{Measuring the dark matter equation of state}
\author[Ana Laura Serra and Mariano Javier L. Dom\'{\i}nguez Romero]{Ana Laura Serra$^{1,2}$\thanks{E-mail:
	serra@ph.unito.it} and Mariano Javier L. Dom\'{\i}nguez Romero$^{3}$\thanks{E-mail:
mardom@oac.uncor.edu} \\
$^1$Dipartimento di Fisica Generale ``Amedeo Avogadro'', Universit\`a degli Studi di Torino, Via P. Giuria 1, I-10125,
 Torino, Italy \\
$^2$Istituto Nazionale di Fisica Nucleare (INFN), Sezione di Torino, Torino, Italy\\
$^3$Instituto de Astronom\'ia Te\'orica y Experimental (IATE), \\
Consejo de Investigaciones Cient\'ificas y T\'ecnicas de la Rep\'ublica Argentina (CONICET),\\
Observatorio Astron\'omico C\'ordoba, Universidad Nacional de C\'ordoba, \\
Laprida 854, X5000BGR, C\'ordoba, Argentina.}

\begin{document}

\date{Accepted XXX Received XXX ; in original form 2011 February 22}

\maketitle
	\begin{abstract}
	The nature of the dominant component of galaxies and clusters remains unknown. While the astrophysics community 
	supports the cold dark matter (CDM) paradigm as a clue factor in the current cosmological model, no direct CDM
	 detections have been performed. \citet{Fab06} have suggested a simple method for 
	measuring the dark matter equation of state that combines kinematic and gravitational lensing data
	 to test the widely adopted assumption of pressureless dark matter.
	Following this formalism, we have measured the dark matter equation of state for first time using improved 
	techniques. We have found that the value of the equation of state parameter is consistent with pressureless dark 
	matter within the errors. Nevertheless, the measured value is lower than expected because typically the masses 
        determined with lensing are larger than those obtained through kinematic methods. 
	 We have tested our techniques using simulations and we have also analyzed possible sources of error that could 
	invalidate or mimic our results. In the light of this result, we can now suggest that the understanding of the 
	nature of dark matter requires a complete general relativistic analysis.
	\end{abstract}

	\begin{keywords}
	equation of state, gravitation, gravitational lensing: strong, gravitational lensing: weak,
	galaxies: clusters: individual, galaxies: kinematics and dynamics, (cosmology:) dark matter
	\end{keywords}

	\section{Introduction}
	Strong evidence, from a large number of independent observations indicates that dark matter is composed by 
	yet unknown weakly interacting elementary particles. Since these particles are required to have small random 
	velocities at early times, they are called cold dark matter (CDM). Many solutions have been proposed 
	to explain its presence but its nature remains obscure. Up to the present the hypothesis of pressureless dark matter
        remains experimentally untested since laboratory experiments have not yield positive results \citep{bertone10}.

	\citet{Fab06} have conceived a novel approach to calculate the density and pressure 
	profiles of the galactic fluid, with no assumptions about their specific form. Such test is based on General Relativity 
	results, the weak field condition, and the probe particle speeds involved (photons and stars). 
	In order to carry out an explicit measurement of the dark matter equation of state (EoS), we have applied this test to 
	galaxy clusters presenting gravitational lensing effects. The advantage of galaxy clusters over galaxies is their larger
	 dark matter concentrations and their vast spectroscopic data, which allows for reliable kinematic profiles.

	\vspace{-0.58cm}
	\section{A relativistic experiment}

	A static spherically symmetric gravitational field is represented by a space-time metric of the form \citep{Misner} $\mbox{d}s^{2}=-c^2 \, e^{2\tilde{\Phi}(r)}\,\mbox{d}t^{2} + \mbox{d}r^{2}/(1-\frac{2 m(r) G}{r c^2}) \, +r^{2} \mbox{d}\Omega^{2}$, where $\tilde{\Phi}(r)=\Phi(r)/c^2$, and $\Phi$ is the gravitational potential.

	Resorting to the Einstein field equations with a consistent static and spherically symmetric 
	stress-energy tensor, and using the mass density definition ($\int 4\pi \rho(r)r^2=m(r)$), the pressure 
	profiles are:

	\begin{eqnarray}
	\nonumber
	\frac{8\pi G}{c^4} p_r(r) \!&=&\! - \frac{2}{r^2}\, 
	\left[ 
	\frac{m(r)G}{c^2 \,r} - r\,\tilde{\Phi}'(r)\left( 1-\frac{2\,m(r)G}{c^2 \, r} \right)  
	\right] ;  \\
	\frac{8\pi G}{c^4}p_t(r) \!&=&\! - \frac{G\,\left[ m'(r)\,r-m(r) \right]}{c^2\,r^3}\, 
	\left[1+r\,\tilde{\Phi}'(r)\right] ..  \\
	\nonumber
	\!&..&\! +\left( 1-\frac{2\,m(r)G}{c^2\,r} \right) 
	\left[ \frac{\tilde{\Phi}'(r)}{r} + \tilde{\Phi}'(r)^2 +\tilde{\Phi}''(r) \right] \, .
	\end{eqnarray}

	$p_r(r)$ and $p_t(r)$ refer to the radial and tangential pressure profiles, which are completely determined by the two functions $\tilde{\Phi}(r)$ and $m(r)$. If these two functions are obtained from  observations, both pressure profiles can be inferred. For a perfect fluid we expect $p=p_t=p_r$.

When analyzing data, it is convenient to assume some simplifying hypothesis. Standard Newtonian physics is obtained in the limit of General Relativity through the following conditions: (i) the gravitational field is weak $\frac{2mG}{c^2r} \ll 1$, $2\Phi \ll c^2$, (ii) the test probe particle speeds are slow compared to the speed of light and (iii) the pressure and matter fluxes are small compared to the mass-energy density. While the first and second conditions are accomplished by galaxies in clusters, photons only fulfill the first condition. The third condition is often applied, since it is related to the nature of the dominant component and to the possibility that this is a pressureless fluid. The novel idea introduced by \citet{Fab06} is to avoid the assumption of the third condition.

Under condition (i) the \emph{tt} component of the Ricci tensor is \citep{Misner} ${\nabla}^2 \Phi \approx -\mathbf{R}_{tt}$, then ${\nabla}^2 \Phi \approx \frac{4\pi\,G}{c^2}(c^2\, \rho + p_r + 2p_t)$. Consequently, the function $\Phi(r)$ can be interpreted as the Newtonian gravitational potential $\Phi_N(r)$ if and only if the fluid is pressureless. In the kinematic regime conditions (i) and (ii) are fulfilled. Due to the second condition, the geodesic equation can be reduced to $\frac{\mbox{d}^2 \mathbf{r}}{\mbox{d} t^2} \approx - {\nabla}\Phi$, where $\mathbf{r}$ is the position vector of the galaxy and $\Phi(r) \neq \Phi_N(r)$. Then, the mass profile obtained from the kinematic analysis is defined by 

\begin{displaymath}
m_{K}(r) = \frac{r^2}{G}\,\Phi_{K}' \approx \frac{4\pi G}{c^2} \int (c^2\rho + p_r + 2p_t)\, r^2\, \mbox{d} r \, ,
\end{displaymath} 
which causes $m_K(r) $ to differ from $ m(r)$. On the other hand, in the case of gravitational lensing, photons are the test particles and condition (ii) is not satisfied. Hence, the geodesic equation needs to be solved exactly to understand the influence of the gravitational field. By applying Fermat's principle and considering an effective refractive index (see \cite{Fab06} for details), the lensing gravitational potential is defined as 

\begin{displaymath}
2\,\Phi_{lens}(r) = \Phi(r) + \int \frac{m(r)}{r^2}\, \mbox{d} r \, ,
\end{displaymath}

where $\mathbf{\nabla}^2 \Phi_{lens}(r) = 4\pi\,\rho_{lens}(r)$. and then $\Phi_{lens}(r) = \int \frac{m_{lens}(r)}{r^2} \, \mbox{d} r \,$. This implies $m_{lens}(r) = \frac{1}{2}\, m_{K}(r) + \frac{1}{2}\, m(r)$. This analysis gives a simple expression for the two functions required to calculate the density and pressure profiles:

\begin{displaymath}
\Phi(r)´=\frac{G m_{k}}{r^2};\,\,m(r)=2m_{lens}(r)-m_{k}.
\end{displaymath} 

It is important to note that a gravitational lensing analysis usually assumes a Newtonian gravitational potential, but in this general case the effective refractive index -physical observable of gravitational lensing- requires a more comprehensive definition of the gravitational potential. 

\vspace{-0.58cm}
\section{Weighing clusters of galaxies}
\label{sec:weight}
Jeans equation offers a direct way to calculate the mass profile via kinematics:

\begin{displaymath}
m_{K}(<r)=-\frac{r\, \sigma_r ^2}{G}\left[\frac{\mbox{d} \ln \rho_{n} }{\mbox{d} \ln r}+\frac{\mbox{d} \ln \sigma_r ^2 }{\mbox{d} \ln r}+2\beta\right ] \, ,
\end{displaymath}

where $m_K(<r)$ is the mass enclosed within a sphere of radius $r$, $\rho_{n}$ is the 3D galaxy number density, $ \sigma_{r} $ is the 3D line-of-sight (l.o.s.) velocity dispersion and $\beta$ is the anisotropy parameter $\beta=1-\langle v_\theta^2+v_\phi^2\rangle/(2\langle v_r^2\rangle)$.
An alternative approach was introduced by \cite{diaferio97}, who suggested the possibility of measuring cluster masses using only redshifts and celestial coordinates of the galaxies. The method they developed, called caustic technique (CT), allows to calculate the mass profile at radii larger than the virial radius, where the assumption of dynamical equilibrium is not valid. On a redshift space diagram (l.o.s. velocity $v$ vs. projected distances $R$ from the cluster center) cluster galaxies are distributed on a characteristic trumpet shape, whose boundaries are called caustics. Since these caustics are related to the l.o.s. component of the escape velocity, they provide a suitable measure of the cluster mass. The CT is a method for determining the caustic amplitude $\cal A(\mbox{r})$, and then the cluster mass profile (see \cite{diaferio97}, and \cite{Diaferio09} for details). 
In order to calculate the kinematic mass profile we have applied the following procedure to a simulated cluster extracted from the Millennium Simulation Run \citep{springel2005}, and  two real rich clusters of galaxies (Coma and CL0024).  The galaxy systems were selected based on four conditions: approximate spherical symmetry, low level of subclustering, gravitational lensing data measurements (only weak lensing reconstruction in the case of Coma) and an important number of measured redshifts of galaxies in order to accomplish our assumptions. Coma and CL0024 have $1119$ and $271$ galaxies within the caustics respectively. 

\begin{figure*}
\includegraphics[scale=.47, bb=60 20 1115 280]{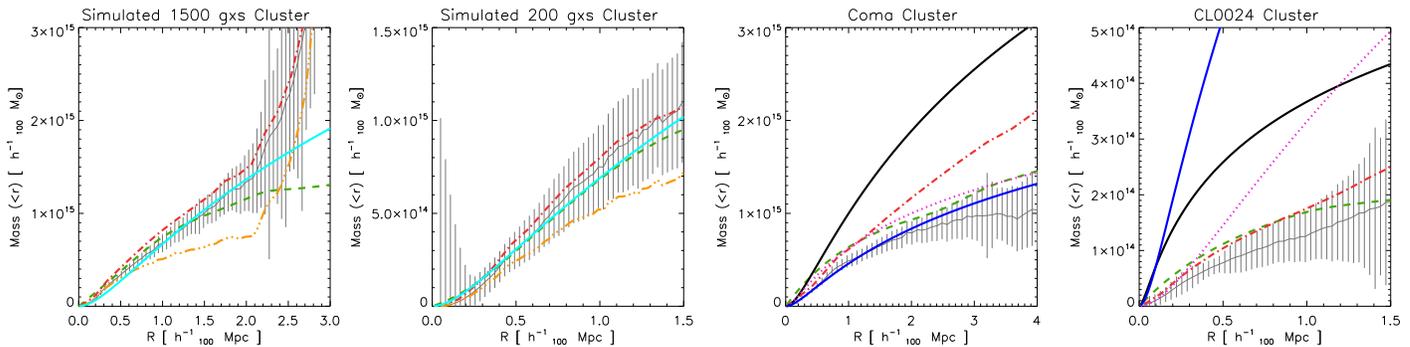} 
\caption{Mass profiles of the simulated (left, with 1000 galaxies and left-center, with 200 galaxies panels) and real clusters (Coma, right-center and CL0024, right panels), for the case 1 (red dash dotted line), 2 (gray continuous line with error bars from bootstrap analysis), 3 (orange dash triple dotted line, only for simulated cluster), from the caustic technique (green dashed line) and from lensing analysis. Lensing profiles are from \citet{Kub07} (black continuous line), \citet{Gava09} (blue continuous line) in Coma and \citet{Kne03} (black continuous line), \citet{Umetsu10} (blue continuous line) in the CL0024. The magenta dotted lines are the NFW kinematic profile from \citet{Lok03} in Coma and the X-ray inferred hydrostatic equilibrium mass profile from \citet{Ota04} in CL0024. In the case of the simulated cluster, the solid cyan line is the NFW fit to the true mass profile, and the kinematic mass profiles were computed selecting a similar number of galaxies than those for the observed clusters, using a magnitude cut-off.}
\label{fig:mass}

\end{figure*}

With the purpose of calculating the mass profile via the Jeans equation, we have used the first steps of the CT to remove interlopers, and an adaptive kernel method (described in \cite{diaferio97}) to estimate the density distribution of galaxies in the redshift diagram. In this way, we are able to obtain a 1D profile for the l.o.s. velocity dispersion $\sigma_R$ and the 2D number density profile $\rho$ (which are simply the second and first moment of the density distribution at each $R$, where $R$ is the projected distance to the center). Using in this novel way the estimated density distribution of galaxies allows us to measure $\sigma_R$ and $\rho$ at several radii, in order to recover the kinematic mass profile with high precision.

In all cases we have used a King profile to fit the number density profile $\rho$ and we have followed the procedure described in \cite{Diaz05} to obtain the 3D number density $\rho_{n}$. In order to determine the velocity dispersion $\sigma_r$, we have applied the Abel inversion technique assuming $\beta=0$. The 2D and 3D profiles are in good agreement with the real profiles of the simulated cluster. This indicates that the assumption of $\beta=0$ is quite adequate. Despite this good agreement, and in order to quantify the impact of $\beta(r)$ on the dark matter equation of state, we have solved the Jeans equation considering three cases: (1) $\beta=0$ and $\beta(r)$ determined by a linear fit to (2) the data of the selected simulated cluster and (3) the data from the most massive clusters of Millennium, up to $1h^{-1}$Mpc. Figure \ref{fig:mass} shows these three mass profiles, together with the caustic mass profile and the true mass profile (NFW best fit to the simulated data). The difference between the upper panels reflects the effect of the number of galaxies tracing the kinematics. To mimic a lensing situation, we have projected the mass of the halos in the field of the simulated cluster and then we have deprojected the 2D density assuming that all the mass belongs to one single halo. The differences between the true  mass profile and the new ``fake'' mass profile can be appreciated only at large radii and their best NFW fits are almost indistinguishable.

\begin{figure}
\centering{
\includegraphics[scale=0.38, bb= 50 40 700 865]{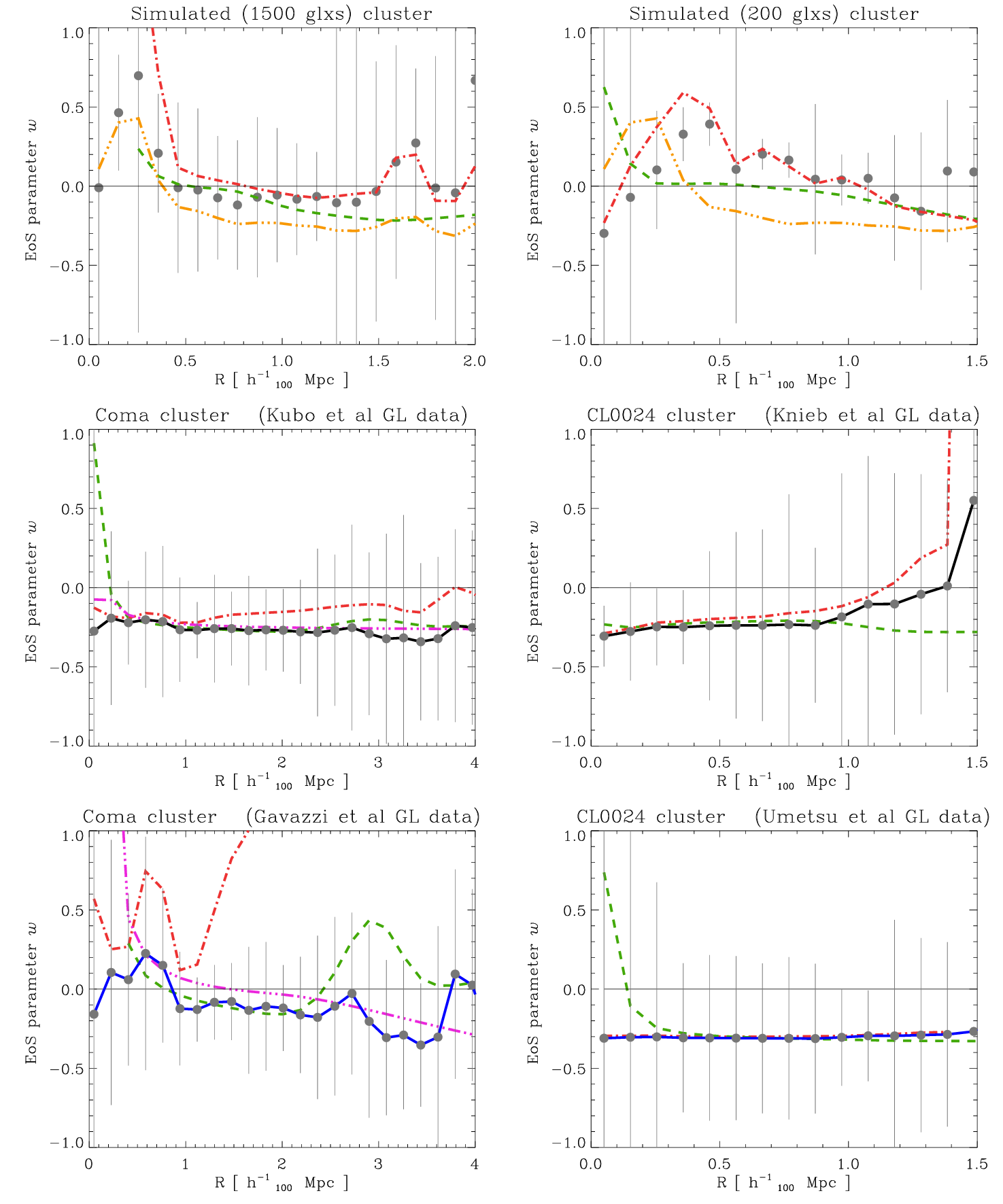} } 
\caption{Dark matter equation of state radial profiles, corresponding to the mass profiles of fig. \ref{fig:mass}, with the same conventions. For clarity we split the EoS parameter determination based on the different mass density profiles obtained from lensing analysis. 
Error bars (at 1 $\sigma$ level) were computed using bootstrap analysis in the galaxies samples used in the determination of the kinematic profile and in the NFW parameters of the dark matter profile inferred from lensing. 
}
\label{fig:state}
\end{figure}

In the calculation of the equation of state, we have combined the tangential and radial pressure, so $w=(p_r+2p_t)/c^2\,3 \rho$. As shown in Figure \ref{fig:mass}, there is a good agreement among the mass profiles of the simulated cluster (except for case 3), implying a null $w$ parameter (Figure \ref{fig:state}). When $\sim 200 $ galaxies are used, the $w$ parameter profiles differ more significantly. This is mainly due to uncertainties of the mass profile and the equation of state parameter determinations which originates from the low number of galaxies of this simulated cluster. The kinematic mass profiles show also some variance due to the presence of inhomogeneities on the radial distributions of galaxies, related with the subclustering.

In Figure \ref{fig:state} can also be seen that, when using the mass profile determined via the caustic technique, $w$ adopts a high positive value in the inner regions of the clusters. This might happen because the CT is very effective in estimating the mass profile in the outskirts but it tends to overestimate it within the virial radius \citep{ser10}.

\vspace{-0.58cm}
\section{Measurement of the dark matter EoS in Coma and CL0024}
The results of the methods described above show a good agreement between the measured profile and the real mass profile in the case of the simulated cluster, independently of the number of galaxies used in the computation of the kinematic mass. Our method is slightly sensitive to the anisotropy parameter; to address this problem, we have calculated the equation of state with the kinematic mass drawn from cases 1,2, and 3 (explained in section \ref{sec:weight}). The resulting profiles, in Figure \ref{fig:state}, show that the anisotropy parameter $\beta$ has a non-negligible impact on the equation of state.
As first result of this work we have found, in the case of the real clusters of galaxies, a good agreement between the kinematic 
mass profiles computed using the caustic technique (heuristic recipe) and the Jeans equation inversion (rigorously correct), but there is a noticeable difference between them and the mass derived from the gravitational-lens model (Figure \ref{fig:mass}, see also \cite{Diaferio09}). 
However, this discrepancy allows us to measure the dark matter equation of state for the first time using clusters of galaxies. 
The resulting equation of state of the dark matter (shown in Figure \ref{fig:state}) behaves as expected when we analyze the simulated cluster, but it adopts an almost constant negative value for the real clusters (however consistent with the Strong Energy Condition) instead of the constant zero value required by CDM. We could attribute that to the anisotropy parameter $\beta$. Nevertheless CT is not strongly dependent on $\beta$, and the equation of state from the caustic profile shows also a trend to be negative. We have checked a number of sources of systematics such as the departure of sphericity and relaxation, the presence of deflecting substructures in the l.o.s. (as those reported in Coma by \cite{adami09}) and the ellipticity of the halo. We have tested the lack of sphericity computing the mass profiles along three different l.o.s., showing no significant differences. The presence of substructures in the l.o.s. has been evaluated not only through the redshift distribution of the cluster galaxies but also deprojecting the 2D mass assuming that the mass of the halos near the l.o.s. belongs to the same cluster (as explained in section \ref{sec:weight}). As for the triaxiality of the halo, we have modified the NFW parameters of the lensing mass, according to the results presented in \cite{corless}, considering an extreme case of axis ratio Q=2.5. None of these tests seems to explain the features we have shown \citep{tfana}, although we stress that a combination of several of them might be responsible for this apparent inconsistency. 

It should be mentioned that the cluster of galaxies CL0024 experienced a merger along the l.o.s. \citep{czoske02} approximately 2-3 Gyr ago. Nevertheless the good agreement between the kinematic mass profile computed and the \cite{Ota04} hydrostatical equilibrium mass in the inner region indicates that the gravitational potential has had time to relax (this was noticed first by \cite{Umetsu10}).
The density profiles determined by lensing methods in CL0024 \citep{Umetsu10,Kne03} are in good agreement with each other. This is not the case of Coma, but it should be recalled that the Coma density profile derived from the weak lensing analysis of \cite{Gava09} was computed in the central region ($R < 1 Mpc$) using a very deep photometry. This density profile was extrapolated to outer radii in our analysis, and the $w$ parameter computed using this profile shows a fair  agreement with the CDM value. The weak lensing profile of \cite{Kub07} span a much wider region using SDSS photometry, and the corresponding $w$ has a constant and negative value.

The EoS parameter recovered using this lensing analysis in Coma shows a similar behavior than those obtained in CL0024 (i.e. a preferred value of $w \sim  -\frac{1}{3}$). There is, however, a trend for the $w$ computed with the lensing profile from \citet{Kne03} to increase towards the external regions.
If further measurements confirm the trend of negative values for the dark matter equation of state $w$, this result could be interpreted in the framework of theories including scalar fields and the possibility of effective negative pressures, or alternative models of gravity. 
It is important to notice that the measured value of $w$ in this work is consistent with the standard pressureless cold dark matter at $1 \sigma$ level. The error analysis uses 30 bootstrap resampling of the galaxies in the computation of the kinematic mass profiles, and  30 dark matter profiles resulting from the errors and degeneracies of the measurements of the NFW halo parameters. The assumption of pressureless dark matter can be further tested by applying the method introduced in this work to a large number of lensing clusters with several galaxy members with measured redshifts. 

\vspace{-0.58cm}
\section*{Acknowledgments}
We acknowledge the constructive comments from the anonymous referee.  We thank Dr. Diego Garc\'{\i}a Lambas, Dr. H\'ector Vucetich, Dr. Osvaldo Moreschi and Dr. Antonaldo Diaferio for many helpful suggestions and comments. 
"The Millennium Run simulation used in this paper was carried out by the Virgo Supercomputing Consortium at the Computing Centre of the Max-Planck Society in Garching. The semi-analytic galaxy catalogue is publicly available at http://www.mpa-garching.mpg.de/galform/agnpaper". This research has made use of the SIMBAD database, operated at CDS, Strasbourg, France.
This work has been partially supported by Consejo de Investigaciones Cient\'{\i}ficas y T\'ecnicas de la Rep\'ublica Argentina (CONICET). Support from the INFN grant PD51 and the PRIN-MIUR-2008 grant ``2008NR3EBK 003'' ``Matter-antimatter asymmetry, dark matter and dark energy in the LHC era'' is gratefully acknowledged. MD also warmly thank to Dr. J. Bass, Dr. A. Valle and Dr. N. G\'andara for their long commitment to public service. 

\label{lastpage}
\vspace{-0.58cm}
\bibliographystyle{mn2e}

\end{document}